\documentstyle[12pt]{article}      
\input epsf.tex
\begin{document}
\newcommand{\ie}{{\it i.e., }}
\newcommand{\bea}{\begin{eqnarray}}
\newcommand{\eea}{\end{eqnarray}}
\newcommand{\be}{\begin{equation}}
\newcommand{\ee}{\end{equation}}
\newcommand{\T}{\bar{T}}
\newcommand{\zb}{\bar{z}}
\newcommand{\mbar}{\bar{m}}
\newcommand{\Sb}{\bar{S}}
\newcommand{\tb}{\bar{t}}
\newcommand{\sbar}{\bar{s}}
\newcommand{\bg}{b^>}
\newcommand{\bl}{b^<}
\newcommand{\G}{{\cal G}}
\newcommand{\ph}{\varphi}
\newcommand{\phb}{\bar{\ph}}
\newcommand{\cg}{c^>}
\newcommand{\cl}{c^<}
\newcommand{\ab}{\bar{a}}
\newcommand{\K}{K\"ahler }
\newcommand{\jb}{\bar{\jmath}}
\newcommand{\Jbar}{\bar{J}}
\newcommand{\Mt}{\widetilde{M}} 
\newcommand{\lef}{\langle} 
\newcommand{\rig}{\rangle} 
\newcommand{\lag}{{\cal L}} 
\newcommand{\vevA}{\lef A \rig}
\newcommand{\veva}{\lef a \rig}
\newcommand{\vevP}{\lef \Phi^i \rig}
\newcommand{\vevw}{\lef W \rig}
\newcommand{\vevp}{\lef \ph^i \rig}

\begin{titlepage}
\begin{center}

\hfill LBNL-39590\\
\hfill UCB-PTH-96/52\\
\hfill hep-th 9611093\\
\hfill \today\\

\vskip 1in

{\Large \bf \mbox{Gaugino Condensation, Loop Corrections} 
\mbox{and 
S-Duality Constraint}}\footnote{Talk given at the NATO ASI
meeting ``Masses of Fundamental Particles", Carg\`ese 96.
This work was supported 
by the Director, Office of Energy 
Research, Office of High Energy and Nuclear Physics, Division of High 
Energy Physics of the U.S. Department of Energy under Contract 
DE-AC03-76SF00098, and the National Science Foundation under grant
PHY-95-14797.\\ 
$\dag$ e-mail: saririan@theorm.lbl.gov }

\vskip 0.7in

Kamran Saririan$^{\dag}$

\vskip 0.2in 

{\sl Theoretical Physics Group \\
Lawrence Berkeley Laboratory \\  and \\ Department of Physics \\
University of California \\ 
Berkeley, CA 94720}\\

\end{center}

\vskip 1.5in

\begin{abstract}

This talk is a brief review of gaugino condensation in superstring
effective field theories and some related issues (such as renormalization of
the gauge coupling in the effective supergravity theories and 
modular anomaly cancellation). As a specific example, we discuss a model containing 
perturbative (1-loop) corrections to the K\"ahler potential and 
approximate S-duality symmetry. 

\vskip 0.3in

\end{abstract}
\end{titlepage}

\newpage

\renewcommand{\thepage}{\roman{page}}
\setcounter{page}{2}
\mbox{ }

\vskip 1in

\begin{center}
{\bf Disclaimer}
\end{center}

\vskip .2in

\begin{scriptsize}
\begin{quotation}
This document was prepared as an account of work sponsored by the United
States Government. While this document is believed to contain correct 
 information, neither the United States Government nor any agency
thereof, nor The Regents of the University of California, nor any of their
employees, makes any warranty, express or implied, or assumes any legal
liability or responsibility for the accuracy, completeness, or usefulness
of any information, apparatus, product, or process disclosed, or represents
that its use would not infringe privately owned rights.  Reference herein
to any specific commercial products process, or service by its trade name,
trademark, manufacturer, or otherwise, does not necessarily constitute or
imply its endorsement, recommendation, or favoring by the United States
Government or any agency thereof, or The Regents of the University of
California.  The views and opinions of authors expressed herein do not
necessarily state or reflect those of the United States Government or any
agency thereof, or The Regents of the University of California.
\end{quotation}
\end{scriptsize}

\vskip 2in

\begin{center}
\begin{small}
{\it Lawrence Berkeley National Laboratory is an equal opportunity employer.}
\end{small}
\end{center}

\newpage
\renewcommand{\thepage}{\arabic{page}}
\setcounter{page}{1}
\setcounter{footnote}{0}



\section*{Introduction}
\indent
 
Amongst the candidates for fundamental unified theories, heterotic
superstring theory  with gauge group $E_8\times E_8$
 seems to be the most promising one.
This is because the spectrum of  the theory easily accomodates the Standard
Model spectrum and gauge structure.  In addition, the underlying gauge group
contains an extra factor of $E_8$ which provides an `hidden sector', which
couples to the observable sector only through gravity,
and, as will be discussed below, plays a crucial role in the mechanism
of supersymmetry breaking.
Furthermore,  the effective theories that
describe the heterotic string in 4 dimensions below the Planck scale
are (nonrenormalizable) locally supersymmetric effective field theories.
Indeed, requiring supersymmetry at energies well above $M_W$ in
order to stabilize the gauge hierarchy, in some sense forces one to
consider locally supersymmetric theories: A unified field theory must include
gravity. Within the framework of General Relativity, a supersymmetric theory
has to be {\sl locally} supersymmetric. This follows from the fact that the
supersymmetry transformation on the metric, or on the vielbein
must include general  coordinate
transformations. Supergravity theories are nonrenormalizable, but can
be consistently viewed as low-energy effective field theories (LEEFT)
for the massless modes of superstring.

A basic feature of superstring constructions in four dimensions is
 the presence of
massless  moduli in the effective field theory.  These fields whose vevs
parameterize the continuously degenerate string vacua, are gauge-singlet
chiral fields; furthermore, they are {\sl exact} flat directions of
the low
energy effective field theory (LEEFT) scalar potential.
Generically, the moduli appear in the couplings
of the LEEFT. For example, the tree level gauge couplings at the string scale
depend on the dilaton, $S$, and the Yukawa couplings as well as the kinetic
terms depend on the $T$-moduli (and $S$ through the \K
potential) . There is mixing of the moduli
beyond tree level, due to both string threshold corrections \cite{dkl} and
field-theoretical loop  effects, as we shall dicuss.

Since the supersymmetric vacua of heterotic strings consist of continuously
degenerate families (to all orders of perturbation theory),
parameterized by the moduli vevs, the latter remain perturbatively
undetermined. This degeneracy can only be lifted by a nonperturbative
mechanism which would induce a nontrivial superpotential for moduli,
and at the same time break supersymmetry.
We shall assume that this nonperturbative mechanism takes place in the
LEEFT and is not intrinsically stringy.
This certainly appears to be the most ``tractible" possibility.
A popular candidate for such a mechanism has been gaugino condensation
which is the focus of this talk. 

As a specific model, we later consider gaugino condensation
in a superstring-inspired effective field theory, with approximate
S-duality invariance \cite{gz,bgsdual} and exact T-modular invariance
(generalization of  the  work in ref. \cite{bgsdual}) and  incorporate an
intermediate scale $M_I$ ($M_{\rm  cond} \ll M_I \ll M_{\rm string}$),
\cite{ks} in order to see how the intermediate-scale threshold corrections
will affect gaugino condensation and supersymmetry breaking. This part
of the talk is based on the work in ref. \cite{ks}.
Incorporating the intermediate-scale threshold corrections into gaugino 
condensation
is non-trivial in the sense that the field-theoretical threshold
corrections at $M_I$ are dilaton-dependent. Hence, these modifications
can have non-trivial implications for supersymmetry breaking by
gaugino condensation. Furthermore, {\it a priori}, nothing
prohibits intermediate scales in the hidden sector.

The outline of this talk is as follows.  In
 the next section, we review  gaugino condensation,
and of duality symmetries (modular and S-duality). 
 We shall discuss our model
in section 3.1, and give the  
 renormalized  \K potential including 1-loop threshold
corrections at an intermediate mass, and constrained by duality symmetries.  
The issues related
to the scalar potential, dilaton run-away, and 
supersymmetry breaking , as well as the role of
the intermediate mass are discussed in section 3.2.  Concluding 
remarks are given in section 4.
Due to the significance of renormalization 
of the field-dependent gauge coupling in such
models and its connection with modular anomaly 
cancellation in the effective theory, we 
give a review of these ideas in the Appendix.  

\section*{Generalities}
\indent 
\subsection*{Gaugino Condensation} 
\indent 

A possible mechanism for breaking supersymmetry within the framework
of ($N=1$, $D=4$) LEEFT of superstring is gaugino condensation in the
hidden sector. In this scenario, the
nonperturbative effects arise from the strong coupling  of  the
asymptotically free gauge interactions at energies  well below $M_{Pl}$.
Corresponding to this strong coupling is the condensation of gaugino
bilinear  $\lef \bar{\lambda}\lambda\rig_{h.s.}$.
Let us briefly remind the reader the overview of the development of gaugino
condensation. It was recognized many years ago that gaugino condensation
in globally supersymmetric Yang-Mills theories without matter does not
break supersymmetry \cite{vy}. In  fact, that dynamical supersymmetry
breaking cannot be  achieved in pure SYM theories was shown by
topological arguments of Witten \cite{witten}.
In the locally supersymmetric case the picture is rather different, namely,
gaugino condensation can break supersymmetry \cite{nilles},
 and the gauge coupling is itself generally field-dependent. When the gauge coupling
becomes strong, it gives rise to gaugino  condensation
at the scale\footnote{These arguments are modified by, for instance, the
requirement of modular invariance
\cite{bgmod}.}
\[ M_{cond}\sim M_{string}\lef {\rm Re}T\rig^{-1/2}e^{-{\rm Re}S/2b_0}  =
M_{string}\lef {\rm Re}T\rig^{-1/2}e^{-1/b_0g_{st}^2},
\] which breaks local
supersymmetry spontaneously
 (\(M_{cond}^3 \sim\lef\bar{\lambda}\lambda\rig_{h.s.} \) ),
and $S$ is the dilaton/axion chiral field.
Supersymmetry breaking in the obesrvable sector is induced
by gravitational interactions which act as `messenger'
between the two otherwise decoupled sectors.

However, there are generally two problems associated with the above scenario.
First, the destabilization of $S$ ---
the only stable minimum of the potential in the
$S$-direction being at  $S\rightarrow \infty$; \ie in the direction where exact
supersymmetry is recovered and the coupling vanishes!  This is contrary to
the expectation that the vacuum is in the  strongly coupled,
confining regime. This problem, the so-called dilaton runaway problem,
is present in most formulations of gaugino
condensation, in particular the so-called `truncated superpotential'
approach \cite{drsw} , where the condensate field is assumed to be much heavier than
the dilaton and therefore is integrated out below $M_{cond}$.
In fact, the dilaton runaway problem is  perhaps a more generic problem
in string phenomenology where the underlying string theory is assumed to
be weakly coupled. We shall return to the dilaton runaway later.

The second difficulty is the large cosmological constant that arises
from the vacuum energy associated with gaugino condensation. An early attempt
to remedy these  difficulties was proposed by Dine {\it et al.}
\cite{drsw}, in the
context of no-scale supergravity
 whereby a constant term, $c\,$,
is introduced in the superpotential which independently breaks supersymmetry
and cancels the cosmological constant. The origin of $c$ is traced to the vev of
the  3 form in 10D supergravity, and is quantized in units of order $M_{pl}$.
Therefore, this approach has the unsatisfactory feature of breaking supersymmetry at the
scale of the fundamental theory.

\subsection*{Duality Symmetries (Modular Invariance and S-Duality)}
\indent

Modular symmetry, with the group $SL(2,{\cal Z})$ subgroup of $SL(2,{\cal R})$
 duality transformations, written in its simplest form:
 \be T\rightarrow \frac{\alpha T -i\beta}{i\gamma T + \delta}, \ee  where
\( \alpha\delta- \beta\gamma=1\) and
$\alpha , \beta , \gamma , \delta$ are integers,\footnote{There is, generally,  one copy of the group per
modulus field $T^i$.} is an exact invariance of the underlying string theory.
However, this symmetry is
anomalous in the LEEFT. Cancellation, or partial cancellation, of
this anomaly in the effective theory can be achieved by
the Green-Schwarz (GS)
mechanism,
which is especially clear in the linear-multiplet formulation of the LEEFT
\cite{anom,anto,gt}. In the corresponding chiral formulation, the
adding of GS counter-terms amounts to modifying  the dilaton \K potential:
\[ \ln (S+\bar{S}) \rightarrow \ln (S+\bar{S}-bG) ,  \]
 where $b=-\frac{2}{3}b_0$, and
$b_0$ is the $E_8$ one-loop $\beta$-function coefficient.
$G=\Sigma_{i}\ln(T^{i}+\bar{T}^{i}-\Sigma|\Phi|^{2})$, and $\Phi$ is any
untwisted sector
 (non-modulus) chiral field in the theory. For simplicity, here we only
consider models where modular anomalies are completely cancelled by GS
mechanism, for example, the (2,2) symmetric abelian orbifolds with no $N=2$
fixed planes, like $Z_3$ or $Z_7$ \cite{anom,anto,gt}.
The role of the
  gauge coupling and its renomalization  in superstring effective theories,
 and the connection with modular anomaly cancellation are reviewed in
 Appendix.
 
 Recently, another type of duality symmetry has been receiving
much attention  in string theories. In this case the
group of duality transformations is $SL(2, {\cal Z})$, but acting on
the field $S$ instead of $T^i$, and is referred to as
$S$-duality. Like its $T$-analogue, this group has a generator
which is the transformation $S\rightarrow 1/S$,
and since $S$ is related to the gauge coupling,
this duality transformation is also referred to as `strong-weak'
duality.  Font {\it et al.} \cite{font} have conjectured that $S$-duality,
like $T$-duality is an exact symmetry of string theory.
More recently, it has been mounting evidence that S-duality is a symmetry of
certain string theories \cite{sduality}. However, these theories
all have $N=4$ or $N=2$ supersymmetries.  At the level of string theory,
there are two different types of S-duality, namely ($i$) those that
map different theories into one another, and ($ii$) those that
map strongly and weakly coupled regimes of the
same theory into each another. Indeed, presently there is no evidence
of an S-dual $N=1$  theory, and it is therefore difficult
to justify the use of S-duality as a true symmetry in the corresponding
LEEFT. However, it has been shown that in the effective
theory, the full $SL(2,{\cal R})$ duality transformation is a
symmetry of the equations of motion of the  gravity, gauge, and dilaton
sector in the limit of weak gauge coupling
\cite{gz,bgsdual}. As in  \cite{bgsdual}, we shall take S-duality as a
guiding principle in constructing the K\"ahler potential for the gaugino
condensate, which is, so far, the least understood element in the
description of the effective theory for gaugino condensation.
That is, we assume that S-duality invariance
is recovered in limit of vanishing gauge coupling,
$S+\bar{S}\,\rightarrow\,\infty$.

\section*{A Specfic Model}
\indent

This model basically generalizes the momel of gaugino condensation
with S-duality of ref. \cite{bgsdual} to the case 
in which  an intermediate scale is present. For details of the 
calculation and a more complete discussion, 
the reader is referred to ref \cite{ks}.

The scheme of generating the intermediate scale considered here does not
involve the spontaneous breaking of the hidden-sector gauge group. Here, we
couple the hidden-sector gauge non-singlet fields $\Phi_i$ to a gauge singlet
$A$. When $A$ dynamically gets a vev, $\Phi_i$ become massive and the
intermediate scale is thus generated. Since $A$ is a singlet, the hidden-sector
gauge group does not break. Such a scheme has interesting implications for
gauge coupling  unification \cite{unif}. 
For consistency, the pattern 
$M_{cond} \ll M_I \ll M_{string}$ is always assumed.
Therefore, we shall integrate out the hidden-matter fields below $M_I$
and the effective lagrangian at $M_{cond}$ will consist of the moduli and
the gauge composites only.

The superpotential for the hidden sector matter fields in our toy model is:
\be
W_{HM}=\frac{1}{2}\lambda^{ij}A\Phi_i\Phi_j +\frac{1}{3} \lambda'A^3 .
\ee
When constructing our model, two symmetry principles have been used to
constrain the Lagrangian: First, the LEEFT must be T-modular invariant to
all orders, according to all-loop string calculations. Second, S-duality
is a symmetry in the weak-coupling limit
$\lef S+\bar{S}\rig \rightarrow \infty$.
We will include the renormalization and intermediate-scale
threshold corrections only in the
dilatonic part of the K\"ahler potential.
We simply write down the \K potential and for the full discussion
we again refer the reader to ref. \cite{ks}.
\be K=-\ln m -3\ln(1-m ^{1/3}Q) + G \ee
where\footnote{Here,  $M_{pl}=1$; and notice that the UV cut-off is taken
to be $M_{string}=(S+\Sb-bG)^{-1/2}$ meaning that the condensation
scale is really in  these units, $Q/(S+\Sb -bG)$.},
\be m=2/g_{eff}^2(M_{cond})
=S+\Sb -bG +3\left[\bg\ln\left(\frac{M_I^2}{M_{string}^2}
\right) +\bl\ln\left(\frac{Q/(S+\Sb-bG)}{M_I^2}\right)\right] ,
\ee 
is the renormalized coupling including the 1-loop threshold corrections 
at the canonically normalized, modular invariant intermediate
mass $M_I$ which can be computed to be:
\be
M_I^2=e^K(K^{\ph\phb})^2|\lambda A|^2=
\frac{|\lambda A|^2e^{G/3}}{9(s+\sbar-bG)}
\left(1 + \frac{b}{s+\sbar-b G}\right)^{-2} . \ee
In these relations, $G$ is the Green-Schwarz term,  
\be G= -3\ln(T+\T - |A|^2) , \ee
and 
$Q=|H|^2e^{G/3}$ (where $H$ is the condensate superfield)
is the modular invariant condensation scale. Various
group theoretical factors are as foolws:
\be
-3b=2b_0=\frac{1}{8\pi^2}C(E_8) ; \hspace{0.3in}
 \bg = (3C_G -C_M)/24\pi^2 , \hspace{0.3in} \bl = C_G/8\pi^2 , \ee
where $C_G$ and $C_M$ are the quadratic Casimirs:
\be C_G=T(adj); \hspace{0.2in} C_M=\sum_{r}n_r T(r);
\hspace{0.2in} T(r)={\rm Tr}_r(T^2), \ee
with $r$ labelling the representations of the gauge group,  
and $n_r$ being the
number of fields in the $r$ representation.

To summarize, our \K potential given in eq. (3) includes the one loop 
renormalization of the dilaton with the intermediate threshold corrections,
as well GS counter terms that ensure modular anomaly cancellation. It is
also constrained by approximate S-duality symmetry as discussed in references
\cite{bgsdual,ks}.

The dynamical fields at the condensation scale in our
model are $S$, $H$, and $T$.
The scalar potential is given by:
\be V=e^K\left[ K^{i\jb}(K_iW+W_i)(K_{\jb}\bar{W}+\bar{W}_{\jb}) - 3|W|^2\right]
\ee
and the \K metric written in terms of $m=2/g_{eff}^2(M_{cond})$ (eq. (5)),
$Q=|H|^2e^{G/3}$, and their derivatives with respect to the scalar fields
is given by:
\bea
K_{i\jb} = m^{-2}\{ m_im_{\jb}\tilde{x} &+& m(\xi-1)m_{i\jb} + (\xi+\xi^2)
(m_iq_{\jb}+m_{\jb}q_i) \nonumber \\
  &+& 3m^2[\xi q_{i\jb}+(\xi+\xi^2)q_iq_{\jb}]  + m^2 G_{i\jb} \}  ,
\eea
where
\[x=m^{1/3}Q, \hspace{0.2in} \xi=\frac{x}{1-x}, \hspace{0.2in}
\tilde{x}=1-2\xi/3+\xi^2/3 ,\] and
\[m_i=\partial_i m , \hspace{0.2in} q= \ln Q ,\hspace{0.2in} q_i=\partial_i\ln
Q
, \hspace{0.2in} etc. \]
\begin{figure}
\begin{center}
\leavevmode
\epsfysize=3.5in \epsfbox{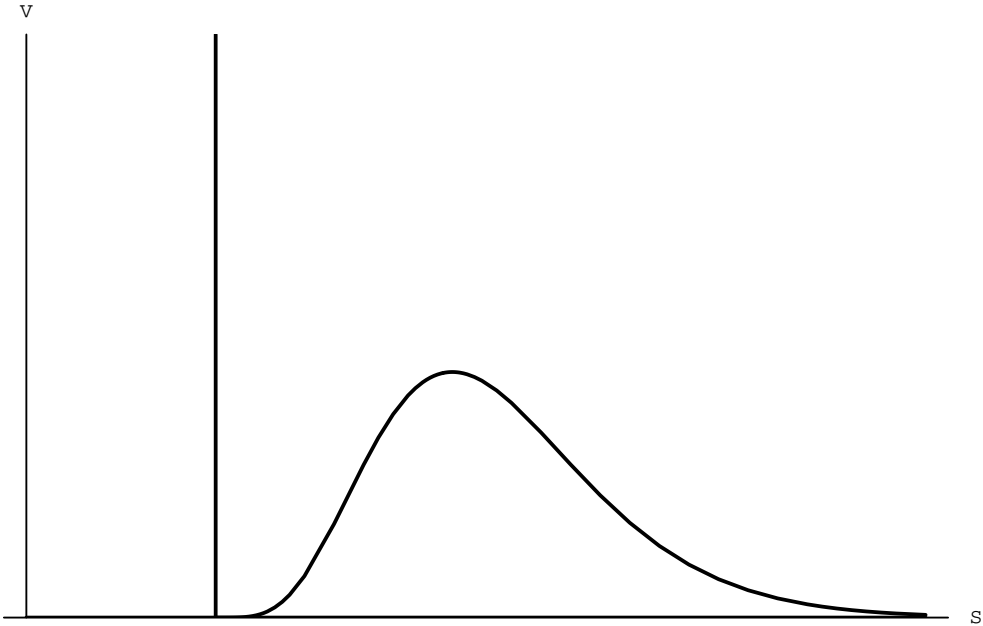}
\caption{The running behaviour of the  dilaton. The minimum on left is at
the boundary  of the kinematically forbidden region.} 
\end{center}
\end{figure}
Notice that $G_{i\jb}=0$ unless $i=j=t$,  $m_{h\jb}=0$, and $q_s=0$.
The  nonperturbative part of the superpotential is of the form
\be W_{NP} = \alpha e^{-S/b}Y^n \left(\ln \frac{Y}{\mu}
\right)^k ,
\hspace{0.2in} Y=He^{S/3\bl}, \ee
with  $n<3$ (the Veneziano-Yankielowicz superpotential is the special case of
$n=3$ and $k=1$). The reason the exponents $n$ and $k$ are introduced
is because it is the
 \K potential (3) that already includes the gaugino condensate
 wave function renormalization,
and so the superpotential should not.

We summarize the results of the numerical computation
and analytic expansions of this model as follows. 
The scalar potential, $V$, is positive semi-definite
and has a nontrivial minimum at finite values of the 
dilaton and the condensate field, at which the following 
relations are satisfied:
\be \vevw\simeq 0, \hspace{0.2in} {\rm and} 
\hspace{0.2in}  m=2/g_{eff}^2(M_{cond})\rightarrow 0 . 
\ee
This is in addition to the usual `runaway' solution at
\(S\rightarrow\infty\) and \(H=0\).
Notice that the second relation tells us that the 
coupling $g_{eff}(M_{cond})$ blows up (at a finite 
value of $S+\bar{S}$). 
 However the nontrivial 
minimum occurs at the boundary of the kinematically
forbidden region of the $(S,H)$ plane. In other words, the 
potential runs in this direction as well! But this the 
correct direction, as the value that it runs to corresponds
to strong coupling at the condensation scale, with a nonzero 
value of the condensate. The similar running behaviours (see Fig. 1)
in both strong and weak coupling directions is attributed to 
S-duality. Notice however that the  relations $\lef V \rig = \lef W \rig = 0$
imply that supersymmetry remains unbroken.

As for the effect of the intermediate mass, the two independent conditions 
$m=\lef W\rig = 0$ imply that \cite{ks}  the parameter $\mu$ of the
nonperturbative superpotential in eq. (11) is `locked' to $M_I$; and that
the parameters of the superpotential which generates $M_I$ allows for a 
phenomenologically sensible hierarchy between the condensation and 
the string scales. 
 
 \section*{Conclusion}
  \indent

We have discussed hidden sector 
gaugino condensation as a possible mechanism for
supersymmetry breaking. In the model which was presented in the 
last section, which included some perturbbative corrections
to the \K potential, as well as a nonperturbative constraint
(S-dulaity), we saw that suprsymmetry remains unbroken. 
Perhaps the most peculiar feature of our model 
is the  running behaviour of the dilaton,
which is schematically
shown in Fig. 1. 
Because the `minimum' on the  left hand side is at the boundary
of the kinematically forbidden region, we hesitate to call
this stabilization of the dilaton. 
The perturbative breakdown of supersymmetry and stabilization of 
moduli of string theory may require the full 1-loop 
corrections to the effective 
supergravity theory  which have been recently calculated \cite{gjs}. 
On the nonperturbative side,
perhaps other stringy nonperturbative effects are more crucial as
pointed ot in ref. \cite{bd}. A realization of
this proposal in the context
 of linear multiplet formulation of gaugino condensate appears in
ref \cite{bgw}. Of course, the exact form of these nonperturbative
corrections are not
yet understood. But one can perhaps expect that
the recent developments in string dualities
can shed some light on the latter, and on the stabilization
of string moduli and supersymmetry breaking.   

\vskip 1cm
\noindent
{\bf Acknowledgements}
\vskip .5cm

I wish to thank Prof. Mary K. Gaillard for useful discussions, and Yi-Yen Wu for 
collaboration on parts of the work that this talk was based on. I also wish
to thank the organizers of Carg\`ese summer school ``Masses of Fundammental 
Perticles" (August 1996).    
This work was supported 
in part by the  Director, Office of Energy Research, 
Office of High Energy and Nuclear Physics, Division of High Energy
Physics of the U.S. Department of Energy under the Contract DE-AC03-76SF00098,
and by the  National Science Foundation under grant PHY-95-14797.

\vspace{0.3in}

\subsection*{Appendix - The Role of the Gauge Coupling}
\indent
\def\theequation{A.\arabic{equation}}


\catcode`\@=11

\setcounter{equation}{0}\indent

In this appendix, we recall a few facts about the perturbative 
corrections of the gauge coupling function in the superstring effective field theories,
as well as the connection with modular invariance of the effective theory.  

As mentioned earlier, in our approach, the one-loop renormalization
of the gauge coupling is completely included in the K\"ahler potential $K$, \ie
the renormalization effects are completely absorbed into $K$ by replacing the
tree-level gauge coupling $S+\bar{S}$ in $K$ by the one-loop renormalized 
gauge coupling. Therefore, it is worthwhile to  discuss
the renormalization of gauge couplings in supestring LEEFTs.   

Let us first recall  the Lagrangian for
supergravity plus super-YM. In the \K covariant formalism \cite{bgg} the 
classical superfield Lagrangian is given by:
\be \lag=\int d^4\theta\{-3E + \frac{E}{8R}f_{ab}(Z)W^{\alpha a}W^b_{\alpha} +
\frac{E}{2R}e^{K/2}W(Z)\} + h.c. ,\ee
where $E={\rm Sdet}E_M\,^{A}$,  $R$ is the curvature scalar  of the 
superspace, and $Z$ stands for the chiral fields in the  theory. 
The first term in eq. (A.1) corresponds to the kinetic energy for the gravity
sector as well as  the chiral fields. The chiral fields enter
through the dependence
of  the spinorial derivatives of $E$ on the \K potential, $K(Z,\bar{Z})$.
The second term describes the super-YM coupling to the theory, with the 
(holomorphic) gauge coupling function $f_{ab}(Z)$ and the YM `field-strength'
superfield 
\[ W_{\alpha} =W_{\alpha a}T^a = (\frac{1}{8} \bar{{\cal D}}^2 -R)e^{-2V}
{\cal D}_{\alpha} e^{2V} ,\]
where $V$ is the vector multiplet containing the YM gauge potential. 
We shall take $f_{ab}=f\delta_{ab}=S\delta_{ab}$ corresponding to the 
bare coupling of the effective superstring theories where $S$ is the
dilaton/axion chiral superfield.  The component form of the second term
contains:
\[ \int d^4x\sqrt{g}\left(-\frac{1}{4} {\rm Re}f \,{\rm Tr}(F^2)-
\frac{1}{4} {\rm Im}f 
\,{\rm Tr}(F\widetilde{F}) \right) , \] 
and thus  Re$f$ is  the YM gauge coupling, while  Im$f$ 
gives the axionic coupling.

Finally in the last term of eq. (A.1), $W(Z)$ is the superpotential which is a
holomorphic function of  the chiral matter fields (independent of  $S$ 
and other internal moduli, until supersymmetry is broken nonperturbatively).   

 In discussing the gauge couplings 
in effective theories,
it is important to to distinguish between the {\sl Wilsonian} couplings, 
 and the physical, or
{\sl effective} couplings. 
In particular in the effective supersymmetric  theories that we  are
considering, there are powerful statements that can be made about the
two types of gauge coupling. The (holomorphic) Wilsonian gauge couplings 
 in supersymmetric YM theories, which appear in the  Wilson effective
action, do not renormalize beyond one loop. These are funcions that appear
in the Wilson effective action, $S_W(\mu)$, the local functional of quantum
operators. In $S_W(\mu)$, only  momenta
between the scale $\mu$ and the UV cut-off contibute to loops.
 The physically measurable
 `effective' (or running) couplings  appear in the c-number valued
generating functional of   1PI  graphs, $\Gamma$; this is in general
a nonlocal functional of background fields that contain the IR momenta
$p<\mu$ running through loops,  as well. Right at the UV cut off,
the Wilsonian couplings, \ie the 
coefficients appearing in front of the operator terms in $S_W$      
are the bare couplings of the theory.  The relation between the 
two effective actions may formally be written as \cite{sv} 
\[ e^{i\Gamma[\Phi_{c\ell},\mu]}=\lef  e^{iS_W[\Phi,\mu]}\rig , \] 
where the expectation value on the right hand side  is taken in the 
the presence of background fields. 
In the supersymmetric YM theories, it is known that, unlike the
Wilsonian  gauge coupling, the effective coupling renormalizes
perturbatively at all orders, and that, indeed, higher order corrections
introduce nonholomorphicities \cite{sv}.   The generalizations of
these results to supergravity effective theories of superstrings 
have been carried out more recently \cite{dkl,gt,kl1,kl2}.

The  gauge coupling in all $N=1$ effective heterotic string constructions
is given at tree level by:
\be g^{-2}_{\alpha}=k_{\alpha}{\rm Re}S=k_{\alpha}g^{-2}_{string}. \ee
 Re$S$ is the `universal' gauge coupling at string scale, and 
$k_{\alpha}$ is the level of the affine Lie algebra associated with the 
factor $G_{\alpha}$ of the product gauge group. Subsequently,
we shall set $k_{\alpha}=1$, and throughout  the analysis $G_{\alpha}$ 
refers to the IR strong group with gaugino condensation. 
The exact Wilsonian coupling is given by the holomorphic
function: $f_W=S+f^{(1)}$, and the moduli dependent one-loop 
(\ie all-loop) correction $f^{(1)}(T^i)$
 has been determined \cite{kl2} (see below).
The effective gauge coupling, with LEEFT-loop corrections to all
orders is given by \cite{kl1,sv}:
\bea
 g_{eff}^{-2} (p^2)={\rm Re}S &+& 
b_0\ln\frac{\Lambda^2}{p^2}
+c K + \frac{T(adj)}{8\pi^2}\ln g_{eff}^{-2} (p^2) \nonumber \\  &-&
\frac{1}{8\pi^2}\sum_r T(r)\ln\det Z^{(r)}_{eff}(p^2)   , 
\eea
where, $b_0\equiv (-3T(adj) +\sum_r n_rT(r))/16\pi^2$ (the
 YM $\beta$-function coefficient),
 and $c\equiv (-T(adj) +\sum_r n_rT(r))/16\pi^2$,
 and $Z$ is the kinetic normalization matrix. 
To one-loop order, one has to evaluate the r.h.s. of the above equation
at tree  level, at 2-loop the r.h.s. is evaluated to one loop, etc.   
The one-loop result has also been obtaind in \cite{gt}.
Threshold corrections due to integrating out the heavy string modes
have been calculated in reference \cite{dkl}. These corrections are 
only dependent on the moduli $T^i$, and not on the dilaton. All the
perturbative dilaton dependences in the effective gauge coupling
arise from {\sl field-theoretical} loop effects.  We have seen in 
section (3) that
threshold corrections in the effective field theory also introduce
dilaton-dependent terms  in the running coupling. 

Let us now turn to the question of modular invariance. As  inputs
from string theory,
for general fields $\Phi^I$ (ignoring for the moment the GS counter terms),
 we have the normalisation matrix for the kinetic term, and the \K
function. The former is given by:
\be Z_{I\Jbar}=\delta_{I\Jbar}\prod_i(T^i+\T^i)^{-q^i_I} + {\cal O}(\Phi^2), \ee
where the rational numbers $q^i_I$ are the modular weights of the field
$\Phi^I$. They depend on the twist sector of the orbifold which gives
rise to  the matter fields $\Phi^I$, and the modulus field $T^i$.
The \K function
at the tree level is given by
\(K=-\ln(S+\Sb)-\sum_i\ln(T^i+\T^i) + {\cal O}(\Phi^2) \). For the  modular
transformation given in eq. (1) of the text, $K$ transforms by the usual
transformation law:
\be K\rightarrow K+F+ \bar{F}, \hspace{0.3in} 
F=\sum_i\ln(i\gamma_iT^i +\delta_i).
\ee  
Under a modular transformation,  the non-modulus chiral fields, 
transforms as:
\be \Phi^I\rightarrow C^I_J(T^i) \Phi^J . \ee
Hence, the kinetic matrix $Z_{I\Jbar}$ transforms according to: 
\be Z\rightarrow (C^\dagger)^{-1}ZC^{-1}. \ee
It follows from eq. (A.5 -- A.7)  
that the reparametrization induced  on  the matter fields by
modular tranformations is given by:
\be C^I_J = C_{0J}^I\prod_i(i\gamma_iT^i +\delta_i)^{q^i_I}, \ee
where $C_{0J}^I$ is moduli independent. 

For a generic supergravity theory with super-YM, under the 
combined transformations:
\(K\rightarrow K+ F+\bar{F}\) and 
\(\Phi^I\rightarrow C^I_J\Phi^J\) with $C^I_J$
holomorphic function of the moduli
$\Phi^I$, the \K invariance of the (exact) integral  of 
the RGE's, \ie eq. (A.3)  imply that:
\be  f_W \rightarrow f_W + cF-\frac{1}{2\pi^2}\sum_r 
 T(r){\rm tr}\ln C^{(r)}, \ee
where $c$ is the group theoretical factor given after  eq. (A.3) above.
For $C^I_J$ and $F$ corresponding to modular transformations, eq's (A.5)
and (A.8), 
this gives:
\be 
 {\rm Re}f_W \rightarrow  {\rm Re}f_W -\frac{1}{16\pi^2}
\sum_i 2 \alpha^i\ln|(i\gamma_iT^i
+\delta_i)|^2 , \ee
with
\be \alpha^i=\sum_I T(\Phi^I)(1-2q_I^i) - T(adj) ; \hspace{0.2in} T(\Phi^I)
=\sum_r{\rm tr}(T^2(r)), \ee
and $T_a(r)$  are the generators of the representations of the 
fields $\Phi^I$. 

Furthermore, the transformation law (A.10) corresponds, up to a modular
invariant function, to the transformation of the logarithm of 
Dedekind function. In fact it will give the complete modular dependent
perturbative correction, $f^{(1)}$ to the Wilsonian 
coupling \cite{dkl,anto,gt} :
\be {\rm Re}f^{(1)}=\delta{\rm Re}f_W=
 -\frac{1}{4\pi^2}\sum_i\alpha^i\ln|\eta(iT^i)|^2 ,\ee 
modulo a moduli independent part which has been argued to be 
a constant in most orbifold models \cite{kl2}. These equations 
are interpreted as 
a parametrization of the string threshold corrections to the gauge
couplings \cite{dkl}.

Modular invariance is restored by including factors of $\eta(iT^i)$
in the superpotential  (see eq (2)), and in the definition 
of the fields,  so as to cancel the 
above modular dependent correction of the gauge coupling,  
as well as by introducing
GS counter term as discussed in the text. 
However, the inclusion of the $\eta$ factors
tends to spoil the boundedness from below  of the scalar potential.
To avoid this, we may restrict ourselves to 
the orbifold models which do not receive string threshold corrections.
These models have been classified \cite{dkl,anto,kl2}. 
For such models, the
modular anomaly is solely cancelled by the GS counter term.

\end{document}